%Paper: 9111048
%From: David Kutasov <dvk@pupthy.Princeton.EDU>
%Date: Sat, 23 Nov 91 16:52:35 est

\input harvmac

\def\np{Nucl. Phys. }
\def\pl{Phys. Lett. }

\def\prl{Phys. Rev. Lett. }

\def\CP{{\cal P}}
\def \CT{{\cal T}}
\def \CO{{\cal O}}
\def \CR{{\cal R}}

\def \CW{{\cal W}}
\def\frac#1#2{{#1\over#2}}
\def\coeff#1#2{{\textstyle{#1\over #2}}}

\def\journal#1&#2(#3){\unskip, \sl #1\ \bf #2 \rm(19#3) }
\def\andjournal#1&#2(#3){\sl #1~\bf #2 \rm (19#3) }

\Title{PUPT--1293, RU--91--49}
{{\vbox {\centerline{Ground Rings and Their Modules in}
\smallskip
\centerline{2D Gravity with $c\le 1$ Matter}
}}}

\centerline{\it David Kutasov}
\smallskip\centerline{Joseph Henry Laboratories,}
\centerline{Princeton University, Princeton, NJ 08544, USA}
\bigskip
\centerline{\it Emil Martinec\footnote{*}
{\rm On leave of absence from Enrico Fermi Inst.\ and Dept.\ of Physics,
University of Chicago, Chicago, IL 60637}
and Nathan Seiberg}
\smallskip
\centerline{Department of Physics and Astronomy}
\centerline{Rutgers University, Piscataway, NJ 08855-0849, USA}
\vskip .2in

\noindent
All solvable two-dimensional quantum gravity models have non-trivial
BRST cohomology with vanishing ghost number.  These states form a ring
and all the other states in the theory fall into modules of this ring.
The relations in the ring and in the modules have a physical
interpretation.  The existence of these rings and modules leads to
nontrivial constraints on the correlation functions and goes a long way
toward solving these theories in the continuum approach.

\Date{11/91}
%\draftmode

Recently, using matrix model techniques, a number of non-critical
string models have been solved exactly
\ref\pail{D. Gross and A. Migdal, Phys. Rev. Lett. {\bf 64}
(1990) 127; M. Douglas and S. Shenker, Nucl. Phys. {\bf B335} (1990) 635;
E. Brezin and V. Kazakov, Phys. Lett. {\bf 236B} (1990) 144.}
\ref\ising{E. Br\'ezin, M. Douglas, V. Kazakov and S. Shenker, Phys.
Lett. {\bf B237} (1990) 43; D. Gross and M. Migdal, Phys. Rev. Lett.
{\bf 64} (1990) 717;  \v C. Crnkovi\'c, P. Ginsparg and G. Moore,
Phys. Lett. {\bf 237B} (1990) 196.}
\ref\bdssgm{T. Banks, M. Douglas, N. Seiberg and S. Shenker,
Phys. Lett. {\bf 238B} (1990) 279; D. Gross and A. Migdal\journal
Nucl. Phys.&B340 (90) 333.}
\ref\douglas{M. Douglas, Phys. Lett. {\bf 238B} (1990) 176.}
\ref\cone
{E. Brezin, V. Kazakov, and Al. Zamolodchikov, Nucl. Phys. {\bf B338}
(1990) 673; D. Gross and N. Miljkovi\'c, Phys. Lett {\bf 238B} (1990) 217;
P. Ginsparg and J. Zinn-Justin, Phys. Lett {\bf 240B} (1990) 333;
G. Parisi, Phys. Lett. {\bf 238B} (1990) 209.}.
Some of these models were shown to be equivalent to certain topological
field theories
\ref\top{E. Witten, Nucl. Phys. {\bf B340} (1990) 281;
R. Dijkgraaf and E. Witten, \np {\bf B342} (1990) 486;
J. Distler, \np {\bf B342} (1990) 523;
E. Verlinde and H. Verlinde, Nucl. Phys. {\bf B348} (1991) 457;
R. Dijkgraaf, E. Verlinde and H. Verlinde, Nucl. Phys.
{\bf B348} (1991) 435.}
and they exhibit unexpected relation to integrable systems.  Despite
some progress in the continuum Liouville description of these theories
\ref\kpz{V. Knizhnik,
A. Polyakov and A. Zamolodchikov, Mod. Phys. Lett {\bf A3} (1988) 819;
F. David, Mod. Phys. Lett. {\bf A3} (1988) 1651;
J. Distler and H. Kawai, Nucl. Phys. {\bf B321} (1989) 509.}
\ref\liounotes{N. Seiberg, Rutgers preprint RU-90-29;
J. Polchinski, Texas preprint UTTG-19-90.}
\ref\corli{M. Goulian and M. Li,
Phys. Rev. Lett. {\bf 66} (1991) 2051.}
\ref\PO{A. Polyakov, Mod. Phys. Lett. {\bf A6} (1991) 635.}
\ref\DFK{P. Di Francesco and D. Kutasov, Phys. Lett.
{\bf261B} (1991) 385; Princeton preprint PUPT-1276 (1991).}
\ref\macros{G. Moore, N. Seiberg and M. Staudacher, Nucl.Phys.
{\bf B362} (1991) 665.},
their surprising integrability is yet to be understood.  In the flat
space version of these theories the null vectors in degenerate Virasoro
representations lead to Ward identities and to the solution of the
theories
\ref\bpz{A. Belavin, A. Polyakov and A. Zamolodchikov, Nucl. Phys.
{\bf B241}, 333 (1984).}.
This fact has led many people to conjecture that the simplicity and
solvability of these quantum gravity models should be associated with
these null vectors.  In this paper we take a step towards a complete
solution of the models using the null vectors.

The first consequence of the existence of degenerate Virasoro
representations in the matter sector of these theories is the appearance
of infinitely many new states in the BRST cohomology
\ref\lzmin{B. Lian and G. Zuckerman\journal Phys. Lett.&254B (91) 417;
Commun. Math. Phys. {\bf 135} (1991) 547.}
\ref\lzcone{B. Lian and G. Zuckerman, Yale preprint YCTP-P18-91.}.
The standard physical fields have the form $\CT=c\bar c \CO
e^{\alpha\phi}$ where $\CO$ is a matter primary field.  In the $(p,p')$
(with $p>p'$) minimal models there are $(p-1)(p'-1)/2$ such fields
\eqn\physmin{\CT_{n,n'}=c\bar c \CO_{n,n'}
e^{[1 +{p \over p'} - {pn'-p'n \over p'}]{\gamma \over 2}\phi} }
($\gamma= \sqrt{2p' \over p}$) labeled by $n=1,...,p-1$ and
$n'=1,...,p'-1$ with $pn'-p'n>0$.  In the (non-compact) $c=1$ theory
there is a continuous set of operators
\eqn\physcon{\CT_{q}=c\bar c e^{iqX/\sqrt{2}} e^{(2 - |q|) {\gamma
\over 2} \phi}  }
($\gamma=\sqrt{2}$) referred to as tachyons, labeled by the momentum $q$
and infinitely many `special states' for integer $q$ labeled by an
integer $s\geq1$
\eqn\physconedis{\CD_{q,s}=c\bar c e^{iqX/\sqrt{2}}\CP_{q,s}(\partial
X,...) \bar\CP_{q,s}(\bar \partial X,...) e^{[2 - (|q|+2s)] {\gamma
\over 2} \phi} }
where $\CP_{q,s}$ is a polynomial in derivatives of $X$ of dimension
$|q|s+s^2$. We will refer to a tachyon with integer momentum
$\CT_{q}=\CD_{q,s=0}$ as a special tachyon.  In the expressions for the
operators \physmin-\physconedis\ we used the bound on the Liouville
exponent of \liounotes.

\bigskip

\centerline{\bf Rings}

In the interesting papers \lzmin\lzcone\ Lian and Zuckerman have shown
that the null vectors in the matter and Liouville representations lead
to more states with other ghost numbers.  In the $c=1$ system, these
have the same $X$ and Liouville momenta as \physconedis\ but have
vanishing ghost numbers (in our convention the physical states have
ghost number one).  In fact, there are three sets of such states with ghost
numbers $(1,0)$, $ (0,1)$ and $(0,0)$.  The states with $(1,0)$ and $
(0,1)$ lead to conserved currents
\ref\witring{E. Witten, IAS preprint IASSNS-HEP-91/51.}
and the $(0,0)$ states lead to a ring \witring.  We will denote these
operators $\CR_{q,s}$.

In the minimal models each highest weight state has two primitive null
vectors, so the ground ring has twice as many elements as the number of
matter primaries.  The vanishing ghost number operators $\CR_{n,n'}$ are
labeled by $n=1,...,p-1$ and $n'=1,...,p'-1$ (without the standard
identification of $(n,n')$ with $(p-n,p'-n')$).  They are given by
polynomials in Virasoro generators of Liouville and matter sectors, as
well as the modes of the ghost number current, acting on
$\exp[-((n-1)+\frac{p}{p'}(n'-1))\frac\gamma2\phi]\CO_{n,n'}$.  Note
that $\CO_{n,n'}\equiv\CO_{p-n,p'-n'}$ but $\CR_{n,n'}
\not\equiv \CR_{p-n,p'-n'}$ due to the different null vectors
used in their construction.  The first of them $\CR_{1,1}$ is the
identity operator.  Explicit construction of some of the other operators
was given in
\ref\expli{P. Bouwknegt, J. McCarthy, and K. Pilch, CERN preprint
CERN-TH.6162/91;  C. Imbimbo, S. Mahapatra, and S. Mukhi,
Tata Institute preprint TIFR/TH/91-27.}
\eqn\onetwoone{\eqalign{
\CR_{2,1}=&|bc-\coeff1{\gamma}(L_{-1}^L-L_{-1}^M)|^2
                e^{-\gamma\phi/2}\CO_{2,1}\cr
        \CR_{1,2}=&|bc-\coeff{\gamma}2(L_{-1}^L-L_{-1}^M)|^2
                e^{-(p/p')\gamma\phi/2}\CO_{1,2}\cr}
}
For example, in pure gravity $(p=3,p'=2)$ $\CR_{2,1}= |bc - {1 \over
\gamma}\partial\phi |^2 e^{-{\gamma \over 2}\phi}$.  There are also
operators with arbitrarily larger negative ghost number.  Note that
unlike the $c=1$ system there are no $(1,0)$ or $(0,1)$
operators and hence there are no conserved currents.

As pointed out by Witten \witring\ the vanishing ghost number operators
are special because they lead to a ring structure.  The ring
multiplication is obtained by considering the operator product expansion
of two vanishing ghost number operators $\CR_m$ and $\CR_{m'}$ in the
BRST cohomology.  Since the product is BRST invariant, and has
vanishing ghost number, it can be written as
\eqn\ringrel{\CR_m(z) \CR_{m'}(w) = \sum _{m''} f_{m,m'}^{m''}
\CR_{m''}(z) +[Q,\CO]   }
for some operator $\CO$ depending on $m$, $m'$, $z$ and $w$.  Here we
have used the fact that $\CR_m(z) $ has dimension zero.  Ignoring the BRST
commutator in \ringrel\ we find a ring with structure constants
$f_{m,m'}^{m''}$.  Below we will examine whether these BRST commutators
can be dropped in correlation functions.

Treating the Liouville field as free Witten \witring\ has shown that in
the non-compact $c=1$ system the ring is generated by
\eqn\ringaplu{a_+=\CR_{1,1}= |bc - {1 \over \gamma}(\partial\phi  - i
\partial X)|^2 e^{-{\gamma \over 2}(\phi -iX)} }
and its conjugate
\eqn\ringaplu{a_-=\CR_{-1,1}= |bc - {1 \over \gamma}(\partial\phi  + i
\partial X)|^2 e^{-{\gamma \over 2}(\phi +iX)} }
i.e. $\CR_{n,s}=a_+^{(|n|+n)/2+s-1}a_-^{(|n|-n)/2+s-1}$ and it has no
relations.  The generators $a_\pm$ have a beautiful interpretation
\witring\ as the phase space coordinates of the free fermions of the
matrix model description of this model, and the whole ring is then
identified as functions on phase space.  Note that the scaling of
$a_\pm$ is $e^{-\gamma\phi/2}$; i.e. they scale like inverse length.
This is precisely the expected scaling of $\lambda$ and its time
derivative $\dot\lambda$.  There is at least one matrix
model operator with these quantum numbers
\ref\ms{G. Moore and N. Seiberg, Rutgers preprint RU-91-29.}
$\frac1{g_{str}}\int dX \psi^\dagger
\lambda^3\psi e^{\pm iX/\sqrt{2}}$ but there may be others.
The power of $\lambda$ in the operator
does not lead to the wrong scaling behavior because of the
factor of the string coupling in the vertex which scales like
the square of the length.

The operators $J_{q,s}$ and $\bar J_{q,s}$ related to $\CR_{q,s}$ with
ghost numbers $(1,0)$ and $(0,1)$, are almost conserved.  Their
divergences $\bar \partial J_{q,s}$ and $\partial \bar J_{q,s}$ are
BRST commutators.  If these commutators can be ignored, these operators
are holomorphic and anti-holomorphic currents \witring\ and lead to a
symmetry $\CW'$.  $\CW'$ is the subalgebra of the algebra $\CW$ of area
preserving diffeomorphisms of the $a_\pm$ plane that preserves the lines
$a_+=0$ and $a_-=0$.  $\CW$ transformations closely related to those of
this symmetry were first noted in the matrix model in
\ref\trans{D. Minic, J. Polchinski, and Z. Yang,
Texas preprint UTTG-16-91; J. Avan and A. Jevicki,
Brown preprint Brown-HET-821 (1991).}.
They were modified and identified as symmetries of the matrix model in
\ms\ where their relation to the special states was also
explained (see also
\ref\wadiaetal{S.R. Das, A. Dhar, G. Mandal and S.R. Wadia,
ETH-TH-91-30.}).
In the continuum approach this symmetry was related to the special
states also in
\ref\klepol{I. Klebanov and A. Polyakov, PUPT-1281 (1991).}.
The elements of $\CW'$ do not have to preserve the lines $a_\pm=0$
pointwise but only as a set. These lines were interpreted in \witring\
as the Fermi surface of the matrix model and $\CW'$ is then the
subalgebra of the matrix model symmetry $\CW$ which is preserved by the
ground state \witring.

We now return to the minimal models.  The matter content of the ground
ring operator $\CR_{n,n'}$ ($\CO_{n,n'}$), and the CFT fusion rules
constrain the multiplication table of the ring.  Assuming that the
Liouville field is free and examining the Liouville momenta of
$\CR_{n,n'}$, it appears that the ring is generated by $\CR_{1,2}$ and
$\CR_{2,1}$
\eqn\minring{ \CR_{n,n'}=\CR_{2,1}^{n-1}\CR_{1,2}^{n'-1} }
(We did not check this expression explicitly in the most general case.)
Unlike the $c=1$ system, $n$ and $n'$ are bounded, and therefore there
must be some relations in the ring.  Examining the Liouville momentum we
conclude $\CR_{1,2}^{p'-1}=0 $ and $\CR_{2,1}^{p-1}=0 $.  It is amusing
to note that the relation $\CR_{1,2}^{p'-1}=0 $ is the relation in the
underlying chiral ring in the LG description of the topological field
theory at the point $p=1$ \top.  It would be interesting to understand
the role of the other relation in that context.

The generator
$\CR_{2,1}$ scales like the eigenvalue of the matrix model (inverse
length) and the other generator $\CR_{1,2}$ scales like the conjugate
momentum (length to the power $-p/p'$).  Motivated by the interpretation
of the ring at $c=1$ and this scaling behavior, we would like to
interpret these generators as the eigenvalue and its conjugate momentum
in the matrix model.  These are precisely the operators $Q$ and $P$ in
Douglas' \douglas\ derivation of the string equation.  We therefore
propose the identification $\CR_{2,1}=Q$ and $\CR_{1,2}=P$ and the
finite ring as functions on this ``phase space.''  Note that this
``phase space'' is not standard because $Q$ and $P$ can be raised only
to finite powers.  It should be pointed out that the matrix model
operators corresponding to $\CR_{2,1}$ and $\CR_{1,2}$ are generally
believed to be given by the fractional powers $Q^{\frac{2p+p'}{p'}}_+$
and $Q^{\frac{p+2p'}{p'}}_+$ of $Q$ and not by $Q$ and $Q^{p/p'}_+$
respectively.  The apparent discrepancy with the scaling properties is
again resolved by recalling the extra factor of the string coupling,
which scales as the $[\frac{(p+p')}{p'}]^{\rm th}$ power of $Q$.  It is
curious that both the $c=1$ and the minimal models have an operator
which scales like inverse length.
Such an operator, $e^{-{1\over 2}\gamma\phi}$, plays a fundamental role
in the Backlund transformation in Liouville theory
\ref\ct{T.L. Curtright and C.B. Thorn, \prl {\bf 48} (1982) 1309;
E. Braaten, T. Curtright and C. Thorn, \pl {\bf 118B}
(1982) 115; Ann. Phys. {\bf 147} (1983) 365;
E. Braaten, T. Curtright, G. Ghandour and C. Thorn, \prl {\bf 51}
(1983) 19; Ann. Phys. {\bf 153} (1984) 147.}
\ref\gn{J.-L. Gervais and A. Neveu, \np
{\bf 199} (1982) 59; {\bf B209} (1982) 125;{\bf B224} (1983) 329;
{\bf 238} (1984) 125; 396; \pl {\bf 151B} (1985) 271;
J.-L. Gervais, LPTENS 89/14; 90/4.}
and provides the relation to its $SL(2,R)$ symmetry \gn\
\ref\tk{F. Smirnoff and L. Taktajan, Univ. of Colorado preprint
(1990).}.

The discussion above generalizes to the fermionic string.  Again, the
BRST cohomology can be analyzed, and when there are degenerate
representations there is nontrivial BRST cohomology at ghost number zero
(and all negative ghost numbers at $\hat c<1$)
\ref\superbrst{P. Bouwknegt, J. McCarthy, and K. Pilch,
CERN-TH-6279-91.},
except that in this case the generating elements are in the Ramond
sector.  For $c<1$ the matter highest weights $\CO_{n,n'}$ are Ramond
for $n-n'$ odd and Neveu-Schwarz for $n-n'$ even.  Hence $\CR_{1,2}$ and
$\CR_{2,1}$ are Ramond operators.  At $c=1$ the special states are easily
constructed using super-$SU(2)$ current algebra as in \witring\ (where
the odd half-integer spin states are in the Ramond sector).
The ring multiplication
table is identical to the bosonic case, the standard $Z_2$ symmetry of
Ramond-Neveu-Schwarz is identical to the $Z_2$ of even vs.\ odd
polynomials in $\lambda$.  Since the generators are Ramond fields, any
representation (see below) contains both Neveu-Schwarz and Ramond
states.

Note that the construction of the ground ring uses very little of the
structure of the theory, simply that it consists of two sectors:
Liouville and matter, and the matter sector has degenerate
representations.  From a null state, general BRST arguments of the type
given by Witten \witring\ predict the existence of BRST cohomology at
ghost number zero.  In fact one can interpret the program of \gn\ for
$c>1$ as a study of this sector of the string Hilbert space.  Indeed,
the Liouville momenta studied there are precisely at the special values
given by the Ka\v c formula.  Clearly, these operators form a closed
operator algebra.  It is crucial that the Liouville exponent for these
fields is always real even for $c>1$.  It is not clear to us why one is
allowed to ignore all the other states in these theories and it remains
to be seen what the physical interpretation of these states is for
$c>1$.

\bigskip

\centerline{\bf Modules}

Now, consider the other operators in the BRST cohomology.  The operators
of fixed ghost number form a module (a representation) of the ring.  To
see that, consider the operator product expansion of $\CR_m$ and an
operator in the cohomology
\eqn\ringmod{\CR_m(z) V_{i}(w) = \sum _{i'} T_{m,i}^{i'}
V_{i'}(z) +[Q,\CO] } where the sum over $i'$ is over the fields in the
cohomology with the same ghost number as $V_i$.  As in \ringrel, we
first ignore the BRST commutator on the right hand side and conclude
that the coefficients $T_{m,i}^{i'}$ represent the ring multiplication.
It is sometimes the case that this representation is not faithful; i.e.
the matrices $T_{m,i}^{i'}$ satisfy more relations than the underlying
ring and represent a quotient of it.

We now examine the various modules which are present in these theories.
We start with the $c=1$ system and consider a tachyon state $\CT_q$ with
generic (not integer) momentum $q$.  An easy free field calculation
shows that for every fractional part of $q$ and every sign of $q$ there
is a separate module.  For $q>0$
\eqn\tachmodp{\eqalign{  a_+ \CT_q &= q^2\CT_{q+1}+[Q,\CO_{q+1}^+]
\cr a_- \CT_q &= 0+[Q,\CO_{q-1}^-] \cr}}
and for $q<0$
\eqn\tachmodm{\eqalign{  a_+ \CT_q &= 0 +[Q,\CO_{q+1}^+]
\cr a_-\CT_q &= q^2\CT_{q-1}+[Q,\CO_{q-1}^-] \cr}}
None of these modules is faithful.  For $q>0$ the ring generator
$a_-$ is represented by zero and for $q<0$ $a_+$ is zero.  This fact has
a simple interpretation in the matrix model.  As explained by Polchinski
\ref\clas{J. Polchinski, Texas preprint UTTG-06-91.},
the tachyons can be thought of as ripples on the Fermi surface.
Therefore, they satisfy the equation of the Fermi surface which for
vanishing cosmological constant are $a_+=0$ for $q<0$ and $a_-=0$ for
$q>0$.

Similarly, the tachyons $\CT_q$ are not in a faithful representation of
the symmetry algebra $\CW'$.  Since the anti-holomorphic part of
$J_{q,s}$ is the anti-holomorphic parts of $\CR_{q,s}=
a_+^{(|q|+q)/2+s-1} a_-^{(|q|-q)/2+s-1}$, only $J_{q,s=1}$ act
non-trivially and even of these, the negative $q$ $J$'s annihilate the
positive momentum tachyons and vice versa.  In terms of the underlying
phase space the interpretation of this fact is interesting.  The $J$'s
generate the algebra $\CW'$ of reparametrizations of the filled Fermi
sea.  It has a subalgebra $\CW''$ of transformations which leave the
Fermi surface invariant pointwise.  Since the tachyons $\CT_q$ ``live''
on the Fermi surface, $\CW''$ acts trivially on them and the tachyons
represent only the quotient $\CW'/\CW''$ which is essentially a Virasoro
algebra.

For integer values of $q$ the relations are different than
\tachmodp\tachmodm.  For $q$ positive
\eqn\stachmod{\eqalign{
a_+ \CT_q &= q^2\CT_{q+1}+[Q,\CO_{q+1}^+]
\cr a_- \CT_q &= \CD_{q+1,s=1} +[Q,\CO_{q-1}^-]
\cr a_+ \CD_{q,s} &= A_{q,s}^+\CD_{q+1,s}+[Q,\CO_{q+1,s}^+]
\cr a_- \CD_{q,s} &= A_{q,s}^-\CD_{q-1,s+1}+[Q,\CO_{q-1,s}^-]  \cr
}}
where $A_{q,s}^\pm$ are calculable coefficients.  Similar relations hold
for $q<0$.  Note that the zero momentum tachyon $\CT_{q=0}$ is
annihilated both by $a_+$ and by $a_-$.  However, the cosmological
constant operator $\phi\CT_{q=0}$ is in the same module with the special
tachyons and the special states.  We conclude that the special states,
the special tachyons and the cosmological constant are all in one
module.  Unlike the tachyon module, here the relation $a_+a_- =0$ is not
satisfied.  This relation was interpreted on the tachyon module as a
consequence of the fact that the tachyons ``live'' on the Fermi surface.
Similarly we would like to argue that since it is not satisfied for the
special tachyons and the special states, these are not ripples on the
Fermi surface.  We conclude that some of the deformations of the
potential cannot be represented as a change in the state of the system.
This observation is consistent with the Minkowski space interpretation
of this theory.  Rotating $X$ to Minkowskian signature, all the states
in the theory are deformations of the Fermi surface \clas.  Indeed, for
Minkowskian $X$ momentum and for macroscoipic Liouville states
there are no special states in the BRST
cohomology.

For $c<1$, the ghost number zero states are at values of $h$ such that
there are null vectors in their Verma modules as well, leading to ghost
number $-1$ BRST cohomology via the same argument that produced the
ground ring.  This structure repeats at each stage, leading to
$(p-1)(p'-1)$ dimensional cohomology at all negative ghost numbers
related to the tower of inclusions of null modules inside one another in
the matter sector \lzmin\ (note that these physical states are not
dressed null states).  The ground ring acts within the ghost number $n$
Hilbert space; thus it is a representation (module) of the ground ring
modulo BRST commutators.  In the fermionic string there is again a tower
of inclusions of null modules, and hence BRST cohomology at every ghost
number.

Each BRST module at negative ghost number is a faithful representation
of the ground ring $\CR$.  There are as many states in the BRST module
as elements of $\CR$, which acts in a nondegenerate way.  This is not so
for the physical state module at ghost number one, which is half the
size due to the identification $\CO_{n,n'}\equiv\CO_{p-n,p'-n'}$.  The
physical state module $\CP$ cannot be a faithful representation and
there must be extra relations defining the action of $\CR$ on $\CP$.

These take the form
\eqn\anoteq{\CR_{1,2}^a\CR_{2,1}^b=0\quad,\qquad a+b=[(p'-1)p/p']\
.}
There are two interesting submodules of this module.  The ring action on
the fields $\CT_{1,n'}$ with $n'=1,...,p'-1$ are annihilated up to BRST
commutators by $\CR_{2,1}$ and satisfy $\CR_{1,2}\CT_{1,n'}=\CT_{1,n'+1}
+[Q,\CO]$.  Similarly, $\CT_{n,p'-1}$ are annihilated up to BRST
commutators by $\CR_{1,2}$ and satisfy
$\CR_{2,1}\CT_{n,p'-1}=\CT_{n-1,p'-1} +[Q,\CO]$ for $1\le n<p-\frac
p{p'}$.  These are analogous to the tachyon modules in the $c=1$ system
\tachmodp\tachmodm.

\bigskip

\centerline{\bf Correlation Functions}

If it is legitimate to drop the BRST commutators in \ringmod\ in
correlation functions, we derive a set of identities
for the amplitudes:
\eqn\notward{< \CR_m V_{i_1}...V_{i_n}>=\sum _{i'} T_{m,i_1}^{i'}
<V_{i'}V_{i_2}...V_{i_n}> = \sum _{i'} T_{m,i_2}^{i'}
<V_{i_1}V_{i'}...V_{i_n}> =...}
Note that these are not Ward identities.  The latter would involve a sum
of $n$ terms in each of which one of the operators in the correlation
function is modified.  Here we have an equality between pairs of
correlation functions.  As we will see, \notward\ is not always
satisfied.  Correspondingly, the BRST commutators in \ringrel\ and
\ringmod\ do not necessarily decouple.  The standard proof of their
decoupling proceeds by moving the BRST charge from the BRST commutator
to all the other operators in the correlation function.  This has the
effect of generating total derivatives on moduli space.  The original
BRST commutator fails to decouple when these total derivatives do not
integrate to zero.  This phenomenon can be equivalently described in
terms of contact terms at the boundaries of moduli space.  It leads to
violations of \notward; the modules are deformed, with the structure
constants $T^i_{jk}$ acquiring dependence on the couplings one turns on
in the action.

Due to the structure of the ring for non-compact $c=1$ described above,
it is enough to consider in this case
\eqn\Aq{A^{(\pm)}(q_1,...,q_n)\equiv\langle a_\pm(z) \CT_{q_1}...
\CT_{q_n}\rangle}
where $\sum q_i\pm1=0$. It is implied in \Aq\ that $n-3$ of the
positions of $\CT_{q_i}$ are integrated over, and the appropriate $\CT$
are stripped of the $c\bar c$ factors in \physcon. The strategy for
extracting information from \Aq\ is to note that on general grounds
$A^{(\pm)}(q_i)$ is independent of $z$; therefore we can compare its
value as $a_\pm$ approaches two different (unintegrated) $\CT_{q_i}$.
This will give a set of relations between different amplitudes \notward.

It is convenient to consider first amplitudes, in which the $\{q_i\}$
satisfy a ``resonance'' condition $\sum(2-|q_i|)=5$. Such amplitudes are
proportional to the volume of space-time and possess integral
representations which have been studied before
\ref\GTW{A. Gupta, S. Trivedi and M. Wise, Nucl. Phys. {\bf B340} (1990)
475.} \PO \DFK.
We will now show that many of their properties are simple consequences
of the action of the ring on the tachyon modules.

Consider the general such correlation function
$\langle\CT_{q_1}..\CT_{q_n}\CT_{p_1}..\CT_{p_m}\rangle$
where $q_i>0$, $p_i<0$. It is known \PO \DFK\ that for $n,m\geq2$
these amplitudes vanish. To derive this fact from the ring we evaluate
\eqn\Anmq{A_{n,m}(q_i,p_j)=
\langle a_+(z)\CT_{q_1}(0)\CT_{p_1}(1)\CT_{q_2}(\infty)
\prod_{i=3}^n\int d^2z_i\CT_{q_i}(z_i)\prod_{j=2}^m\int d^2w_i
\CT_{p_j}(w_j)\rangle}
in two different limits. As $z\rightarrow0$ we can replace
$a_+\CT_{q_1}$ by $\CT_{q_1+1}$ using \tachmodp. It is important that
the BRST commutator in \tachmodp\ does not contribute to \Anmq.
Commuting $Q$ to $\CT_{q_i}$ ($\CT_{p_j}$) we find a total derivative in
$z_i$ ($w_j$).  One can show that it integrates to zero; indeed, it is
readily verified that near all boundaries of moduli space the integrand
goes to zero in an appropriate region in momentum space (and is
analytically continued to vanish everywhere else). Hence, as
$z\rightarrow0$ we find
\eqn\Aone{A_{n,m}(q_i,p_j)=q_1^2  \langle \CT_{q_1+1}
\prod_{i=2}^n\CT_{q_i}\prod_{j=1}^m  \CT_{p_j}\rangle}
On the other hand, as $z\rightarrow 1$, we use
\tachmodm\ (again, one can explicitly verify that the boundary terms
due to the BRST commutators vanish) and conclude that
$A_{n,m}(q_i,p_j)=0$ (for $n,m\geq2$). Comparing to \Aone\ we find the
desired result.

The cases $n=1$ (any $m$), and $m=1$ (any $n$) have to be discussed
separately:

\noindent{}1) $m=1$: In this case, momentum conservation and the
``resonance'' condition enforce $p=p_1=-(n-2)$. As we saw before, for
integer $p_1$ \tachmodm\ should be replaced by \stachmod\ i.e. acting
with $a_+$ produces one of the special states \physconedis. Although one
can proceed this way, a much more useful relation is obtained by
exchanging $\CT_{p_1}\leftrightarrow \CT_{q_2}$, such that \Anmq\ takes
the form
\eqn\nusha{A_{n,1}(q_i)=
\langle a_+(z)\CT_{q_1}(0)\CT_{q_2}(1)\CT_{p_1}(\infty)
\prod_{i=3}^n\int d^2z_i\CT_{q_i}(z_i)  \rangle}
In this case it is easy to see that we can use the naive
form of \notward\ to find
\eqn\anotnus{
q_1^2\langle \CT_{q_1+1}\CT_{q_2}\prod_{i=3}^n\CT_{q_i}\CT_p\rangle=
q_2^2\langle \CT_{q_1}\CT_{q_2+1}\prod_{i=3}^n\CT_{q_i}\CT_p\rangle }
Redefining $\CT_q={\Gamma(1-|q|)\over\Gamma(|q|)}\tilde\CT_q$, we conclude
that
\eqn\F{F(q_1,...q_n)=\langle\tilde\CT_{q_1}...\tilde
\CT_{q_n}\tilde\CT_p\rangle}
is periodic in all its arguments (subject to the constraint $\sum_{i=1}^n
q_i=n-1$):
\eqn\period{F(q_1+1,q_2,...)=F(q_1, q_2+1,...)=...}
An explicit evaluation \DFK\ yields $F(q_i)={\rm const}$, but one can
not determine the periodic function $F$ from the action of the ring. The
algebraic reason for this ambiguity is that the theory has a number of
different modules.  The relation of the ring cannot determine the
``reduced matrix elements'' of different modules.

\noindent{}2) $n=1$: In this case we have to be careful with the BRST
commutators in \tachmodp, \tachmodm. As $z\rightarrow0$, one can use
\tachmodp\ naively; hence, $A_{1,m}=\langle \CT_{q_1+1}\prod_{i=1}^m
\CT_{p_i}\rangle$. On the other hand, as $z\rightarrow 1$, we find a
BRST commutator which does not decouple.  The point is that since $q_1$
is fixed kinematically ($q_1=m-2$), an on shell tachyon arises in the
channel where all $w_i$ simultaneously approach zero.  This can be shown
to lead to a finite boundary contribution of the appropriate total
derivative. Hence here $a_+\CT_p\not=0$ $(p<0)$.  For a quantitative
analysis it is more convenient to replace $a_+$ by $a_-$ in \Anmq, and
imitate the procedure of the first case.

To emphasize the ambiguity of the ring relations \notward\ in $c=1$ by a
periodic function \period, it is useful to consider the open $c=1$
string theory on the disk
\ref\BK{M. Bershadsky and D. Kutasov, Princeton preprint PUPT-1283 (1991),
Phys. Lett. {\bf B}, in press.}. The qualitative considerations
used above are valid there as well. Equation \tachmodp\ takes the form
(for $q>0$)
\eqn\opentp{\eqalign{a_+\CT_q=&q\CT_{q+1}+[Q,V^+]\cr
                     a_-\CT_q=&[Q,V^-]}}
and similarly for $q<0$. The analysis of which correlation functions
vanish is quite different in this case; its conclusions are in agreement
with \BK. The case $\langle a_+\prod_{i=1}^n \CT_{q_i}\CT_p\rangle$ with
$q_i>0$, $p<0$ leads, as above, to
\eqn\Sopen{\langle\prod_{i=1}^n\CT_{q_i}\CT_p\rangle=
  \prod_{i=1}^n{1\over\Gamma(q_i)}
  G(q_1,...,q_n)}
where $G$ is a periodic function of the $q_i$. $G$ is actually a
complicated function of momenta \BK:
$G(q_i)=\prod_{l=1}^{n-1}{1\over\sin\pi(q_1+q_2+...+q_l)}$.  It does not
seem to be obtainable from the action of the ring.

So far we have only discussed ``resonant'' amplitudes in which $\mu$ is
in a sense zero.  In generic amplitudes (``finite $\mu$'') the situation
is more involved.  The BRST commutator terms in \ringmod\ - \tachmodm\
cannot be ignored.  One can still study the deformations of the Fermi
surface in the presence of tachyon perturbations.  As conjectured in
\witring, the equation $a_+a_-=0$ should be modified for non-zero $\mu$
to $a_+a_-=\mu$.  (Note that this relation is not an operator relation
in the theory.)  Unlike \witring, from our point of view the relation
$a_+a_-=0$ is obtained as a relation in the tachyon module
\tachmodp\tachmodm.  Following \witring, we conjecture that it is also
modified to $a_+a_-=\mu$.  Indeed, one can show (using the methods of
\DFK) that the three tachyon amplitude with generic momenta $q_i$
satisfies
\eqn\threeta{<(a_+a_-  -\mu)\tilde\CT_{q_1}\tilde\CT_{q_2}\tilde
\CT_{q_3}>=0}
In the presence of more tachyons the operator $(a_+a_- -\mu)$ does not
vanish.  For example
\eqn\fourta{<(a_+a_-
-\mu)\tilde\CT_{q_1}\tilde\CT_{q_2}\tilde\CT_{q_3}\tilde\CT_{q_4}>
=\mu^{{1\over 2} \sum |q_i| -1}}
and more complicated expressions for higher $n$ point functions.  This
fact has an obvious interpretation in the spirit of \clas\ and \witring.
In the presence of more tachyons the Fermi surface is deformed and no
longer satisfies $a_+a_ -=\mu$.  As explained after equation \notward,
from the world-sheet point of view, this deformation can be understood
as a contact term leading to non-zero correlation functions for the BRST
commutators in \ringmod\ - \tachmodm.

To summarize, the main point in this note is the importance of the
relations in the ring
and in its non-faithful modules.  These relations constrain the
correlation functions.  However, in order to fully utilize the ring and
its relations, we have to get better control of the contact terms at the
boundaries of moduli space.  In the infinite radius $c=1$ system the
tachyons are small deformations of the Fermi surface in the matrix
model, and therefore, the relations in the tachyon module have a natural
interpretation as determining the location of the Fermi surface.  The
contact terms should therefore be associated with the deformation of the
Fermi surface due to the presence of other tachyons.  We expect them to
appear as multi tachyon states in the right hand side of $a_\pm
\CT_q$.  An important open problem is to explicitly determine these
contact terms.

In the minimal models the ring and its relations should be more powerful
than in the $c=1$ system.  The ambiguity in the correlation functions in
the $c=1$ theory stems from the existence of infinitely many modules and
the relations in the ring cannot determine the ``reduced matrix
elements.''  In the minimal models the physical, ghost number one fields
are all in one module and therefore a similar ambiguity is not present.
Unfortunately, for these theories we do not have an interpretation of
the relations in the module analogous to the Fermi surface at $c=1$.  An
interesting relation $\CR_{1,2}\CR_{2,1}=0$ in the submodules of
$\CT_{1,n'}$ and $\CT_{n,p'-1}$ is similar to the equation $a_+a_-=0$ in
the tachyon module.  An amusing possibility is that this relation will
be deformed to $\CR_{1,2}\CR_{2,1}=g_{str}$ which is reminiscent of the
tree level string equation $Q_0P_0=g_{str}$ in terms of the constant
(zeroth order in derivatives) terms in the KdV operators $P$ and $Q$.
Since we know that the string equation is analytic in the matrix model
coupling constants, $t_k$, and that these are analytic in the conformal
field theory couplings \macros, we expect that the string equation can
be computed perturbatively in these couplings using free field
techniques.  The non-analyticity of the solution will arise only from
the solution of this equation.  We hope that a better understanding of
this issue will lead to the entire KdV structure and the Virasoro and
$W$ constraints of these theories and will make the connection of
Liouville theory to the matrix model and to topological field theory
complete.

{\bf Note added:} After the completion of this work we learned that I.
Klebanov and A. Polyakov had obtained some of our $c=1$ results using
another approach and that P. Bouwknegt, J. McCarthy and K. Pilch had
independently found the ground ring in the $\hat c \le 1$ fermionic
system.

It is a pleasure to thank T. Banks, M. Douglas, B. Lian, G. Moore, A.
Polyakov, S. Shenker, C. Vafa, H. Verlinde, E. Witten, A.B.
Zamolodchikov and G. Zuckerman for useful discussions.  This work was
supported in part by DOE grants DE-FG05-90ER40559, DE-AC02-76ER-03072
and DE-AC02-80ER-10587 and an NSF Presidential Young Investigator Award.

\listrefs
\bye